# MODELLING OF Pb RELEASE DURING PORTLAND CEMENT ALTERATION

Anne Bénard[a], Jérôme Rose[b], Jean-Louis Hazemann[c], Olivier Proux[d], Laurent Trotignon[e], Daniel Borschneck[b], André Nonat[f], Laurent Chateau[g], Jean-Yves Bottero[b]

[a]present address INERIS, Institut National de l'Environnement Industriel et des Risques, INERIS-Méditerrannée, Europôle de l'Arbois, BP 33, 13545 Aix En Provence Cedex 04, France, anne.benard@ineris.fr, phone : 33 (0) 4 42 97 14 85, fax : 33 (0) 4 42 97 14 89

[b]CEREGE, Centre Européen de Recherche et d'Enseignement en Géosciences de l'Environnement, UMR 6635 (CNRS-Aix –Marseille University), FR 3098 ECCOREV, Europôle de l'Arbois, BP 45, 13545 Aix En Provence Cedex 04

[c]Laboratoire de Cristallographie, BP 166, 38042 Grenoble Cedex 09, France

[d]LGIT (UMR CNRS-Université Joseph Fourier) 1381 rue de la piscine, Domaine Universitaire, 38400 St Martin d'Hères, France.

[e]CEA de Cadarache, Commissariat à l'Energie Atomique, DTN/SMTM, Laboratoire de Modélisation des Transferts dans l'Environnement, Bât. 307, 13108 Saint-lez-Durance, France

[f]Université de Bourgogne-CNRS, Laboratoire de Recherche sur la Réactivité des Solides, UMR 5613, Faculté des Sciences Mirande, BP 47870, 21078 Dijon Cedex, France

[g]ADEME, Agence De l'Environnement et la Maîtrise d'Energie, 2 square Lafayette, BP 406, 49004 Angers, France

**5232 words**
**9 figures**
**3 tables**

**Abstract**

Complex cementitious matrices undergo weathering with environmental exchange and can release metallic pollutants during alteration. The molecular mechanisms responsible for metal release are difficult to identify, though this is necessary if such processes are to be controlled. The present study determines and models the molecular mechanisms of Pb release during Portland cement leaching. Since Pb release is strongly related to its speciation (i.e. atomic environment and nature of bearing phases), the first objective of the present work is to investigate the evolution of Pb retention sites together with the evolution of the cement mineralogy during leaching. Complementary and efficient investigation tools were used, namely XRD, µ-XRF and XAFS. The second goal is to reproduce our results with a reactive transport code (CHESS/HYTEC) in order to test the proposed speciation model of Pb.

Combined results indicate that both in the unaltered core and altered layer of the leached cement, Pb(II) would be retained through C-S-H "nano-structure", probably linked to a $Q_1$ or $Q_{2P}$ silicate tetrahedra. Moreover in the altered layer, the presence of Fe atoms in the atomic environment of Pb is highly probable. Unfortunately little is known about Fe phases in cement, making the interpretation difficult. Can Fe-substituted hydrogranet ($C_3AH_6$) be responsible for Pb retention? Modelling results are consistent with Pb retention through C-S-H in layers and also in an additional, possibly Fe-containing, Pb-retention phase in the altered layer.

## 1. Introduction

Industrial cements may contain trace amounts of metals. In many cases (bridges, pipes…), concretes are in permanent contact with water and are likely to be altered and therefore to release metals into the environment. To assess environmental risks, i.e. long-term evolution of the solids, it is necessary to develop a strategy that combines the determination of the structural and crystal-chemistry evolution of the solid matrix down to the molecular level (when possible) to a modelling approach of leaching experiments at macroscopic scale (laboratory and field tests). The validation with mechanistic models of short-term leaching experiments is a pre-requisite for long-term prediction.

Lead is one heavy metal of particular interest due to its well-known deleterious effects on human health and its presence as a trace element in cement and treated waste matrices. Even though the mechanisms of lead fixation in some cement minerals have been investigated, i.e. using pure phases [1-6] the evolution of lead speciation within Portland cement submitted to leaching remains hypothetical.

This study aims at i) understanding lead behaviour within cement down to the molecular level and during leaching ii) testing the proposed Pb speciation model by integrated reactive transport simulation of the sample evolution. Numerical calculations are generally used to approach release prediction. Among the modelling techniques available in the literature, reactive transport codes, like CHESS-HYTEC [7, 8] have already demonstrated their efficiency in modelling cement matrix leaching.

Two kinds of approaches of cement leaching modelling are found in the literature: modelling the release of chemical elements into eluates (majors and minors) [9] and modelling the mineralogical evolution in the solid matrix itself during leaching [10-12]. The second approach is used here even though it is more difficult. Among the difficulties of such an approach, the lack of thermodynamic data, especially for trace metal fixation mechanisms in a complex matrix such as a Portland cement, is of particular concern. In the case of a trace element such as Pb, it is crucial to determine its speciation with regard to both the atomic environment and the sorbing phase.

The first part of the present study aims at understanding the evolution of Pb speciation during the alteration of Portland cement. As the Pb concentration is very low and the cement mineralogy complex, different complementary and investigation tools were used (X-Ray Diffraction (XRD), micro-X-Ray Fluorescence (µ-XRF), X-ray Absorption Near Edge Structures (XANES) and Extended X-ray Absorption Fine Structures (EXAFS)).

Interpreted in terms of chemical reactions, results were added to the thermodynamic database of the reactive transport simulation code (CHESS/HYTEC) to reproduce Pb behaviour in the solid matrix during alteration and to test the speciation model of Pb.

## 2. Materials and method

### 2.1. Materials

The Portland cement characteristics are presented in Table 1. The cement was hydrated with demineralised water, with a water/cement ratio of 0.5 during 28 days. Because of EXAFS detection limits, 2000 mg/kg of doping Pb salts, as ($Pb(NO_3)_2$), were added into the hydration water. This Pb concentration is still below the solubility limit of Pb oxide or hydroxide.

### 2.2. Leaching experiments

The cement sample was leached for 42 days using a dynamic leaching system: the CTG-Leachcrete [13]. The sample is immersed into the leaching cell, initially filled with demineralised water. Chemical unbalance between cement and water remains constant because of a demineralised water turnover using a process of water vaporisation and condensation (0.5 ml./hour). This system favours a dynamic leaching of the solid. The total volume of the solution in contact with the cement was 1490 ml for a cement mass of 264 g.

To accelerate alteration, the temperature of the leaching cell was maintained at 40°C and water acidity at pH 5, using $HNO_3$, 0.1M. The experiment was performed under ambient atmosphere. Solutions were regularly sampled and analysed. An ICP-AES spectrometer (Jobin-Yvon) was used for chemical analyses.

**2.3 Analytical techniques**

The evolution of cement mineralogy with leaching time was analyzed using XRD by comparison of the unaltered core and the altered layer. The diffractometer used was a Philips PW3710 type, with a Co source. Samples were analysed from 5° to 70°, with a 0.02° step and record time of 12 seconds per step. The unaltered core and the altered layer of the leached sample were separated and analysed individually.

On a polished section of the leached cement, profiles of chemical elements were performed from the unaltered core to the altered layer using µ-XRF with a resolution of 40 µm (Eagle µProbe). The cement sample was first embedded in a 3M Scothcast Electrical Resin #3. No chemical treatment was performed on the sample. Profiles were obtained after integration of the signal measured on a 9000*8000 (x*z) µm area, with a step of 144 µm and record time of 15 seconds per point.

The Pb atomic environment was studied using EXAFS experiments performed at the Pb $L_{III}$-edge on the former beamline CRG-BM 32 at ESRF (Grenoble, France). The fluorescence mode was used and the yield was measured by 30 Ge-solid-state detectors (Canbera). Eight to ten scans were recorded and integrated to increase the signal to noise ratio. The unaltered core and the altered layer were separated and ground into fine powders for analyses. No chemical treatment was performed on the sample. Some reference compounds were also analysed in transmission mode, at NSLS on the X-23A2 beamline (Brookhaven, New York, USA) for calibration purposes. Lead EXAFS spectra for both cement layers and references were treated using a software suite developed by Michalowicz [14].

Modelling of the Pb release was performed with the CHESS-HYTEC v3.2 chemical-transport calculation code [8]. This code considers porosity evolution, as minerals dissolve or precipitate during leaching [11]. Porosity and diffusion coefficient were linked using the Archie Law [15]. The truncated Davies formula was used to evaluate activities in the CHESS code. HYTEC parameters were the same as laboratory experimental leaching conditions (40°C, pH 5, 50 days), and calculations were made in one space dimension. The alteration rate was defined in order to obtain a depth of portlandite dissolution similar to the experimental portlandite dissolution depth, and therefore a diffusion coefficient of $10^{-10}$ $m^2/s$ was initially considered. An 18000 µm- sample depth was simulated and 2 mesh grid sizes were considered along this profile: 100 µm in the first 3000 µm from the leachate-cement interface and 200µm in the last 15000µm. The CHESS database was used, except for the minerals mentioned in Table 2. C-S-H solubility was described using the model developed at 25°C by [16]. For both external and internal surfaces of C-S-H particles, this model considers the formation of silicate tetrahedra chains on the CaO planes by insertion of a bridging silicate tetrahedron; also, deprotonation and complexation of calcium on silanol sites, on both external and internal surfaces (interlayer). Formation constants at 40°C were calculated using the Van't Hoff law.

## 3. Results and discussion

### 3.1. Mineralogical evolution of the cement matrix after leaching: experimental results

Fig. 1 compares XRD results for unaltered core and altered layer (from 0 to 1500 µm) and Fig. 2 shows intensity profiles of the major elements present in cement matrix. In Fig. 2 and 3A, fluorescence intensities were divided by the corresponding mean intensity for profile comparison. The altered layer at 1300-1400 µm depth is characterised by partial dissolution of portlandite and calcium monosulfoaluminate; secondary ettringite precipitates from 900 to 1300-1400 µm. The altered layer also contains hydrotalcite and calcite. C-S-H is still present in the altered layer but may be replaced by a siliceous gel in the first 200 µm from the interface leachate/cement [10, 17]. Hydrogarnet ($C_3AH_6$) is also present in the altered layer from 1000 to 2000 µm. Results are in agreement with previous studies [10, 17]. According to the literature, C-S-H would be decalcificated in the altered layer [18, 19]. Fe is hardly released and accumulates in the altered layer [17, 18] (Fig 2, 3A).

### 3.2 Pb behaviour during leaching

Eluates analyses indicated a very low release of Pb (0.058% of Pb initial concentration). The Pb profile (Fig. 3A), reveals a significant increase of Pb relative content in the altered layer from 1500 to 100 µm from the surface, then a decrease in the first 100 µm from the leachate/cement interface. Comparison of the µ-XRF profiles of Pb with major elements constitutive of cement mineralogy (Fig. 3A), suggests similar trends between Pb and Si in both the unaltered core and ettringitic zone. The affinity of Pb for C-S-H, already demonstrated in literature for unaltered cement or individual cement minerals [5, 6, 17], can explain the Pb-Si co-location. In the surface layer, from 0 to 700 µm from the surface, it is worth noting a strong correlation between Pb and Fe (Fig. 3B) while the Pb-Si correlation decreases. This result indicates that iron rich phases may play a role in Pb fixation in addition to C-S-H. From micro-XRF elemental profiles C-S-H and iron rich phases are suspected as Pb bearing minerals so Pb-O, Pb-Ca, Pb-Si and Pb-Fe were the selected atomic waves for numerical simulation of EXAFS spectra. All theoretical amplitude and phase functions were tested and calibrated on $PbNO_3$, $Ca_2PbO_4$, $PbSiO_3$ and $Pb[Fe_3(SO_4)_2(OH)_6]_2$ EXAFS spectra [20]. Pb-Pb atomic interactions, present in Pb hydroxide, were also considered and tested on $Ca_2PbO_4$, $PbSiO_3$ EXAFS spectra.

### 3.2. Evolution of Pb retention site with leaching

Pb(II) is the only stable redox state under Eh and pH conditions of Portland cement interstitial solution [21]. This oxidation state was confirmed by studying Pb $L_{III}$ XANES (results presented in [20]). XANES spectra also pointed out that, in both unaltered core and altered layer, $PbNO_3$, $PbCO_3$, $PbSO_4$ are not present, meaning that $Pb(NO_3)_3$ initially introduced was completely dissolved and that compounds like cerussite or hydrocerussite were not present (Fig. 4).

The Pb $L_{III}$ edge EXAFS spectra of unaltered core and altered layer are compared in Fig. 5 and simulation results, in a radius of 4Å around Pb atoms, are detailed in Table 3. A slight difference on EXAFS spectra as well as on Radial Distribution Function -RDF - (Fig. 5A and 5B) could be observed. Simulations of Pb $L_{III}$ EXAFS spectra in the unaltered core indicate that 3.6 oxygen atoms at 2.24Å, 0.3 silicon atoms at 3.31 Å, and 1.1 calcium atoms at 3.68 Å surround Pb (Table 3). In the case of the altered layer, two EXAFS models lead to solutions that are mathematically equivalent. For both calculations a water molecule was added to improve simulation. The first fit indicates that Si and Ca are present in the second coordination sphere and results indicate 3.2 and 0.5 O atoms at respectively 2.24 and 2.94 Å, 0.5 Ca atoms at 3.70 Å and 0.3 Si atoms at 3.36 Å. With the second fit, Si and Fe are also present in the second coordination sphere and results indicate 3.2 and 0.5 O atoms at respectively 2.24 and 2.93 Å, 0.7 Fe atoms at 3.67 Å and 0.4 Si atoms at 3.37 Å. The second solution gives a better reproduction of the experimental curves (Fig. 6) and suggests the presence of iron in the atomic environment of Pb. The signal to noise ratio of the EXAFS spectra does not permit the discrimination between both atomic environments and a combination of the solutions could also be considered. Micro-XRF indicating the co-location of iron and Pb in the surface layer (Fig. 3B) suggests that Fe could be present in the Pb atomic environment. Moreover the association of both elements exists in other matrices and iron oxi-hydroxides are known to strongly fix lead.

The EXAFS and micro-XRF data on the Si-Pb relations confirms previous studies indicating the strong affinity between Pb and C-S-H [1-6]. Both in the unaltered core and the altered layer, part of the Pb atoms could be retained by C-S-H, probably linked to a $Q_1$ or a $Q_{2P}$ silicate [5, 6] (Fig.7).

As discussed above, the presence of iron in the atomic environment of Pb in the altered layer is highly probable but requires further evidence and investigation. If the presence of iron in the second coordination sphere of Pb in the altered layer is confirmed, ferric phases would be involved in the fixation of lead. Unfortunately the accurate identification of iron phases involved in Pb fixation as well as the fixation mode faces a major problem. Indeed the exact nature of iron rich phases in cement is poorly known especially in the altered zone. It is generally assumed that the minerals formed during the hydration of $C_4AF$ are similar to those formed from tricalcium aluminate ($C_3A$) [22-26]. Iron can replace Al in AFm and AFt phases and in hydrogarnet as recently shown [27]. Moreover amorphous FeOOH phases were formed [27]. But the exact status of iron in cement paste is more difficult to address than during the hydration of pure $C_4AF$. One main question concerns the possible presence of iron in C-S-H in the altered layer.

The present study clearly shows the presence of hydrogarnet in the altered layer. But what is the exact role of such Fe substituted mineral? At this stage of the study, it is impossible to conclude on Pb potential retention by hydrogarnet, as no Pb-Fe distances are available for this kind of retention. To study the potential retention of Pb by oxy-(hydroxide), the Pb-Fe distances in the cement altered layer were compared with Pb-Fe distances adsorbed on goethite (FeOOH) [28] and amorphous iron oxide (HFO or $Fe_2O_3.nH_2O$, n=1-3) [29] at Pb $L_{III}$-edge. In goethite, Pb-Fe distances vary from 3.30 to 3.36 Å and from 3.80 to 3.93 Å [28], and with HFO the Pb-Fe distance ranges from 3.29 to 3.36 Å which are quite different from the 3.67 Å distance in cement. Other Pb-Fe distances available from the literature, like in magnetoplumbite ($Pb(Fe,Mn)_{12}O_{19}$) [30] were also compared to the results of the present study. In magnetoplumbite, both 3.63 Å and 3.36 Å distances occur but in the present case, the 3.36 Å distances was not found. Comparison with literature indicates that it is not possible to conclude the exact atomic arrangement of Pb within ferric phases in the altered layer of the cement.

Pb retention by C-S-H and a ferric phase in the altered layer are sufficient to explain the low release of Pb, as these phases are resistant to leaching.

### 3.3 Modelling of cement alteration and Pb behaviour after leaching

Because of the uncertainties on Pb retention in iron containing phases in the altered layer, only Pb retention through C-S-H was considered for modelling. To translate Pb retention through C-S-H, Pointeau retention isotherms [19] were used and associated with the C-S-H model developed by Nonat et al. [16]. Reactions 1 and 2 represent Pb linked to $Q_1$ and $Q_{2P}$ sites, respectively, at 25°C.

$$SiOH + Ca^{2+} + Pb^{2+} + 3H_2O - 4H^+ = SiOCaPb(OH)_3$$

$$\log K(25°C) = -33.4 \quad (1)$$

$$SiOH + SiOH + SiO_{2(aq)} + Ca^{2+} + Pb^{2+} - H_2O - 4H^+ = SiOH\text{-}CaSiOPb\text{-}SiOH$$

$$\log K(25°C) = -23.3 \quad (2)$$

Fig.8 shows CHESS-HYTEC simulation results for 50 days leaching calculation in CTG-leachcrete conditions. The portlandite dissolution front delimits the altered layer corresponding to 1200 µm from the interface leachate/cement. The secondary ettringite precipitation front is also reproduced from 400 to 700 µm. C-S-H and hydrotalcite are dissolved only in the first 200 µm. Porosity is also reported and increases due to portlandite dissolution, then ettringite dissolution and finally C-S-H and hydrotalcite dissolution. This model reproduces well the dissolution and precipitation phenomena. However, the model must be improved to predict the depth of dissolution and precipitation fronts.

Fig.9 shows the simulated Pb profile, expressed in concentration, normalised by the mean corresponding concentration. Pb accumulation in the first 200µm is well reproduced, corroborating C-S-H role in Pb retention. However, Pb normalised intensity of µ-XRF profile is 60% higher than Pb modelled normalised concentration. Only Pb sorption to non-decalcified C-S-H was considered, but is not sufficient to reproduce experimental data. One explanation can be related to retention of Pb in iron phases. Indeed the role of the C-S-H in Pb retention leads to an underestimate of the relative Pb increase in the altered layer, perhaps due to Pb fixation to iron rich phase. The underestimation of the Pb content in the altered layer can be considered as indirect proof of the role of the iron rich phase(s).

Even if improvements are necessary for a rigorous reproduction of experimental data, the model reproduced all the phenomena observed in experimental data. This result shows the efficiency of the methodology for modelling the behaviour of low metal concentration in complex matrix and allows considering a first validation of the model.

## 3. Conclusion

This study aimed at introducing the methodology used to model Pb behaviour during a Portland cement artificial weathering, using a reactive transport simulation tool (CHESS-HYTEC). Precise thermodynamic data, especially on Pb speciation evolution during leaching, are necessary to achieve an efficient model. µ-XRF and XAFS experiments gave information on Pb atomic environment and the results were associated with the evolution of the cement matrix with leaching (µ-XRF, XRD). Experiments were carried out on cement doped with lead and results indicated a low release of Pb, which accumulates in the altered layer during leaching. According to EXAFS data, both in unaltered core and altered layer, one fraction of Pb would be retained by C-S-H, one of the most resistant phases during leaching. In the altered layer, the presence of Fe in the Pb atomic environment is highly probable. Iron oxi-(hydroxide) or cerussite/hydrocerussite would not retain Pb, and a structure like a Fe substituted-hydrogarnet is a candidate phase for Pb. But at the stage of the study, no definitive conclusion can be reached.

Pb retention by C-S-H in both unaltered core and altered layer was expressed using appropriate chemical reactions and associated thermodynamic data for modelling. The resulting model reproduces the general trend of experimental laboratory profiles, even if improvements could be done using thermodynamic data on Pb retention through ferric phases in the altered layer.

The methodology applied to understand Pb released during cement alteration, and then to model this behaviour is efficient and necessary in term of long-term behaviour prediction problems. It should be applied in the case of other complex matrices used for example in waste valorisation processes. This program is currently carried out in ARDEVIE laboratory (Aix-En-Provence, France). This laboratory is a collaboration between INERIS, CEREGE and SMA (Syndicat Mixte de l'Arbois), partners of the project.

**Acknowledgements**

This work was performed by CEREGE. The authors thank ADEME, ATILH, SCORI for technical and financial support, CEA of Cadarache for modelling and ESRF team for XAFS analyses. CHESS and HYTEC development is supported by the Pôle Géochimie Transport (Armines, CEA, EDF, IRSN, Lafarge, Total). We also thank Annette Johnson for her contribution.

Table 1

Chemical and mineralogical composition of Portland cement

| Chemical composition | $SiO_2$ | $Al_2O_3$ | $Fe_2O_3$ | CaO | MgO | $SO_3$ | $K_2O$ | $Na_2O$ |
|---|---|---|---|---|---|---|---|---|
| (%) | 20.68 | 4.78 | 2.89 | 63.17 | 3.60 | 2.75 | 0.57 | 0.20 |

Table 2

Thermodynamic data used for modelling, logK is given at two temperatures. LogK of portlandite, ettringite, calcium monosulfoaluminate, hydrogranet and hydrotalcite are from [31, 32], logK at 85°C of hydrogarnet from [33], Hydrotalcite from [34] and C-S-H from [16].

| Minerals | Reactions | Log K |
|---|---|---|
| Portlandite | $Ca(OH)_2 = Ca^{2+} + 2H_2O - 2H^+$ | **25°C**: -22.79; **40°C**: -21.78 |
| Ettringite | $(CaO)_6(Al_2O_3)(SO_3)_3(H_2O)_{32} = 2Al^{3+} + 3SO_4^{2-} + 6Ca^{2+} + 38H_2O - 12H^+$ | **25°C**: -56.11; **40°C**: -53.06 |
| Calcium monosulfo aluminate | $(CaO)_4(Al_2O_3)(SO_3)(H_2O)_{12} = 2Al^{3+} + SO_4^{2-} + 4Ca^{2+} + 24H_2O – 12H_2O$ | **25°C**: -70.92; **40°C**: -67.04 |
| Hydrogarnet ($C_3AH_6$) | $(CaO)_3(Al_2O_3)(H_2O)_6 = 2Al^{3+} + 3Ca^{2+} + 12H_2O - 12H^+$ | **25°C**: -79.2; **85°C**: -65.2 |
| Hydrotalcite | $(MgO)_4(Al_2O_3)(H_2O)_{10} = 4Mg^{2+} + 2Al^{3+} + 17H_2O - 14H^+$ | **25°C**: 73.75 |
| C-S-H particle | $Ca_2Si_2O_5(OH)_2 + 4H^+ + H_2O = 2Ca^{2+} + 2H_4SiO_4$ | **25°C**: -29.6; **40°C**: -28.21 |
| C-S-H internal and external surface sites | $SiOH = SiO- + H^+$ | **25°C**: -11.8; **40°C**: -11.8 |
| | $SiOH + Ca^{2+} = SiO\text{-}Ca^+ + H^+$ | **25°C**: -9; **40°C**: -9.14 |
| | $SiOH + SiOH + Ca^{2+} = SiOCaOSi + 2H^+$ | **25°C**: -20.4; **40°C**: -19.51 |
| | $SiOH + Ca^{2+} + H_2O = SiOCaOH + 2H^+$ | **25°C**: -24.85; **40°C**: -23.77 |
| | $SiOH + SiOH + SiO_{2(aq)} = SiOH\text{-}HOSiOH\text{-}SiOH + 2H_2O$ | **25°C**: 7; **40°C**: 6.81 |
| | $SiOH + SiOH + SiO_{2(aq)} = SiOH\text{-}HOSiO^-\text{-}SiOH + 2H_2O + H^+$ | **25°C**: -4.8; **40°C**: -4.99 |
| | $SiOH + SiOH + SiO_{2(aq)} + Ca^{2+} = SiOH\text{-}HOSiCa^+\text{-}SiOH + 2H_2O + H^+$ | **25°C**: -2; **40°C**: -2.33 |
| | $SiOH + SiOH + SiO_{2(aq)} + Ca^{2+} = SiOH\text{-}HOSiCaOH\text{-}SiOH + H_2O + 2H^+$ | **25°C**: -17.85; **40°C**: -16.96 |
| | $SiOH + SiOH + SiO_{2(aq)} = SiOH\text{-}^-OSiO^-\text{-}SiOH + 2H_2O + 2H^+$ | **25°C**: -16.6; **40°C**: -16.79 |
| | $SiOH + SiOH + SiO_{2(aq)} + 2Ca^{2+} = SiOH\text{-}Ca^+SiCa^+\text{-}SiOH + 2H_2O + 2H^+$ | **25°C**: -11; **40°C**: -11.47 |
| | $SiOH + SiOH + SiO_{2(aq)} + 2Ca^{2+} = SiOH\text{-}HOCaSiCaOH\text{-}SiOH + 4H^+$ | **25°C**: -42.7; **40°C**: -40.73 |
| | $SiOH + SiOH + SiO_{2(aq)} + Ca^{2+} = SiOH\text{-}^-OSiCaOH\text{-}SiOH + H_2O + 3H^+$ | **25°C**: -29.65; **40°C**: -28.76 |
| | $SiOH + SiOH + SiO_{2(aq)} + 2Ca^{2+} = SiOH\text{-}Ca^+SiCaOH\text{-}SiOH + H_2O + 3H^+$ | **25°C**: -26.85; **40°C**: -26.10 |

Table 3

Results of numeric simulation for Pb $L_{III}$ edge EXAFS spectra in the altered and unaltered layers of the leached Portland cement, with

Q residue = sum of $[k^3. (k.khi(k)_{experimental} - k. khi(k)_{theoric})^2 / (k^3.(k. khi(k)_{experimental}))]^2$,

$khi^2$ residue = $(N_{ind} - N_{par})/(N_{pt} - N_{par}) \sum^{kNpt}_{k1} [(k.\chi(k)_{experimental} - k.\chi(k)_{theoric}) / \delta(k)]^2$, with $N_{ind}$: independant parameters number; $N_{par}$: parameters number; $N_{pt}$: points number; $\delta(k)$: standard deviation of $k\chi(k)$

| Simulated spectra | Atoms nature | Atoms number (N) | Distance R(Å) | Q residue | $khi^2$ residue |
|---|---|---|---|---|---|
| **Unaltered core** | | | | | |
| From 1.09 to 3.78 Å | O | 3.6 ± 0.7 | 2.24 ± 0.03 | 8.64e$^{-3}$ | 3.66e$^{-3}$ |
| | Si | 0.3 ± 0.1 | 3.31 ± 0.06 | | |
| | Ca | 1.1 ± 0.2 | 3.68 ± 0.02 | | |
| **Altered layer (first solution)** | | | | | |
| From 1.12 to 3.83 Å | O | 3.2 ± 0.7 | 2.24 ± 0.03 | 8.84e$^{-3}$ | 3.18e$^{-3}$ |
| | O | 0.5 ± 0.1 | 2.94 ± 0.03 | | |
| | Si | 0.3 ± 0.1 | 3.36 ± 0.06 | | |
| | Ca | 0.5 ± 0.1 | 3.70 ± 0.02 | | |
| **Altered layer (second solution)** | | | | | |
| From 1.12 to 3.83 Å | O | 3.2 ± 0.6 | 2.24 ± 0.03 | 6.78e$^{-3}$ | 2.44e$^{-3}$ |
| | O | 0.5 ± 0.1 | 2.93 ± 0.03 | | |
| | Si | 0.4 ± 0.1 | 3.37 ± 0.06 | | |
| | Fe | 0.7 ± 0.1 | 3.67 ± 0.03 | | |

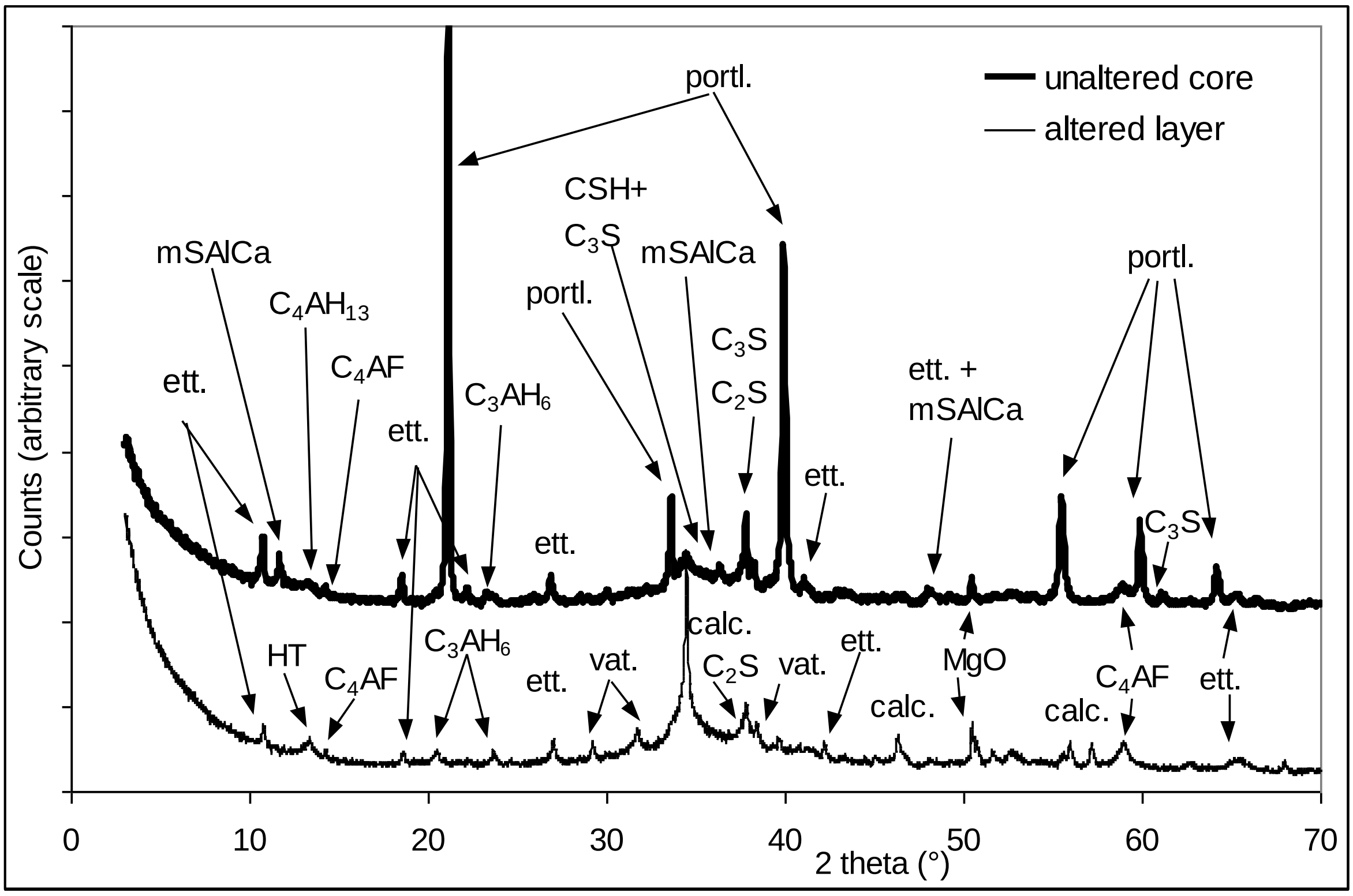

Figure 1. XRD spectra of unaltered core and altered layer of the leached Portland cement. Ett: ettringite, mSAlCa: calcium monosulfoaluminate, HT: hydrotalcite, portl.: portlandite, calc.: calcite, vat.: vaterite

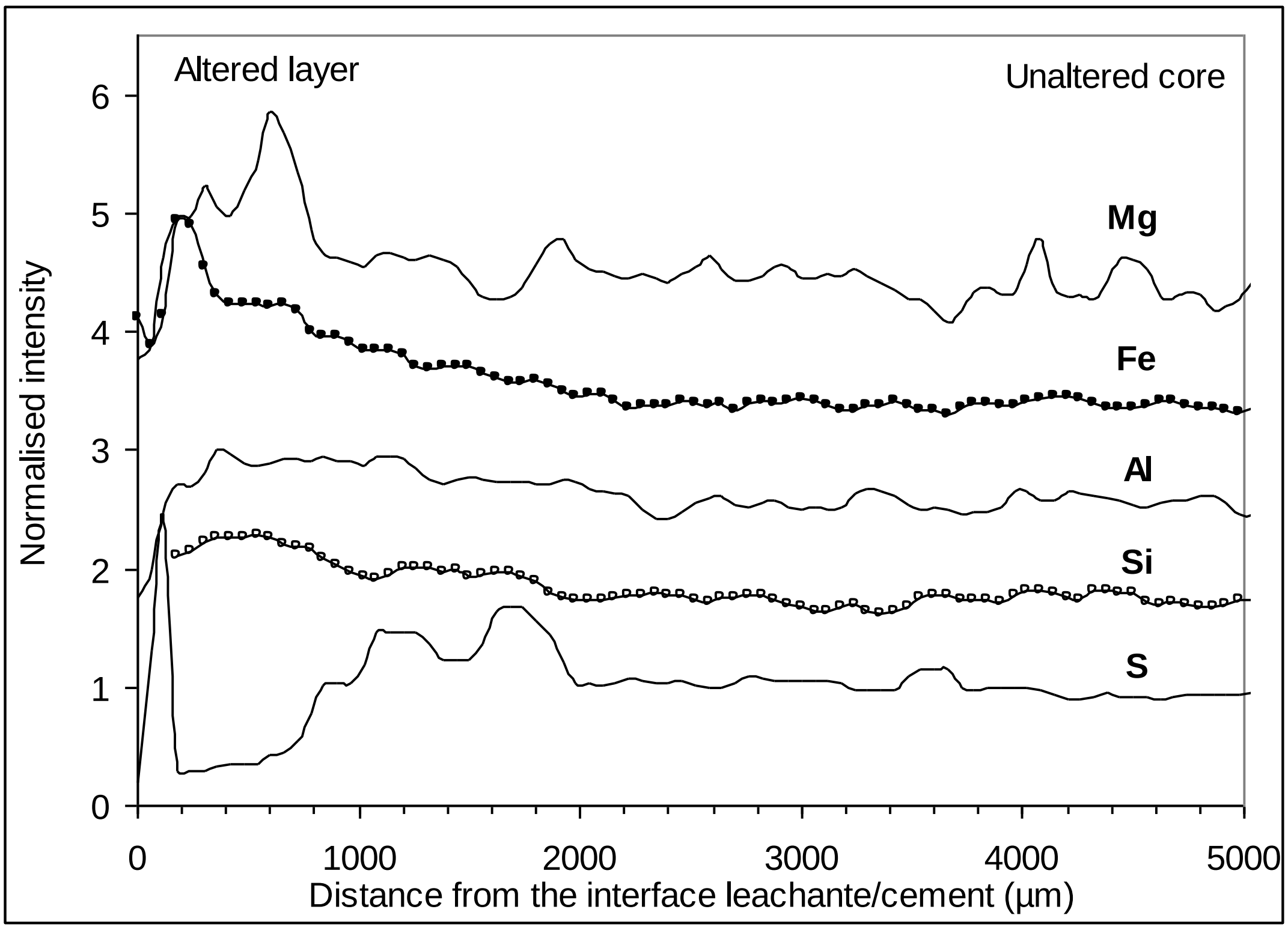

Figure 2. µ-XRF profiles of major elements, from the unaltered core to the altered layer of the leached Portland cement.

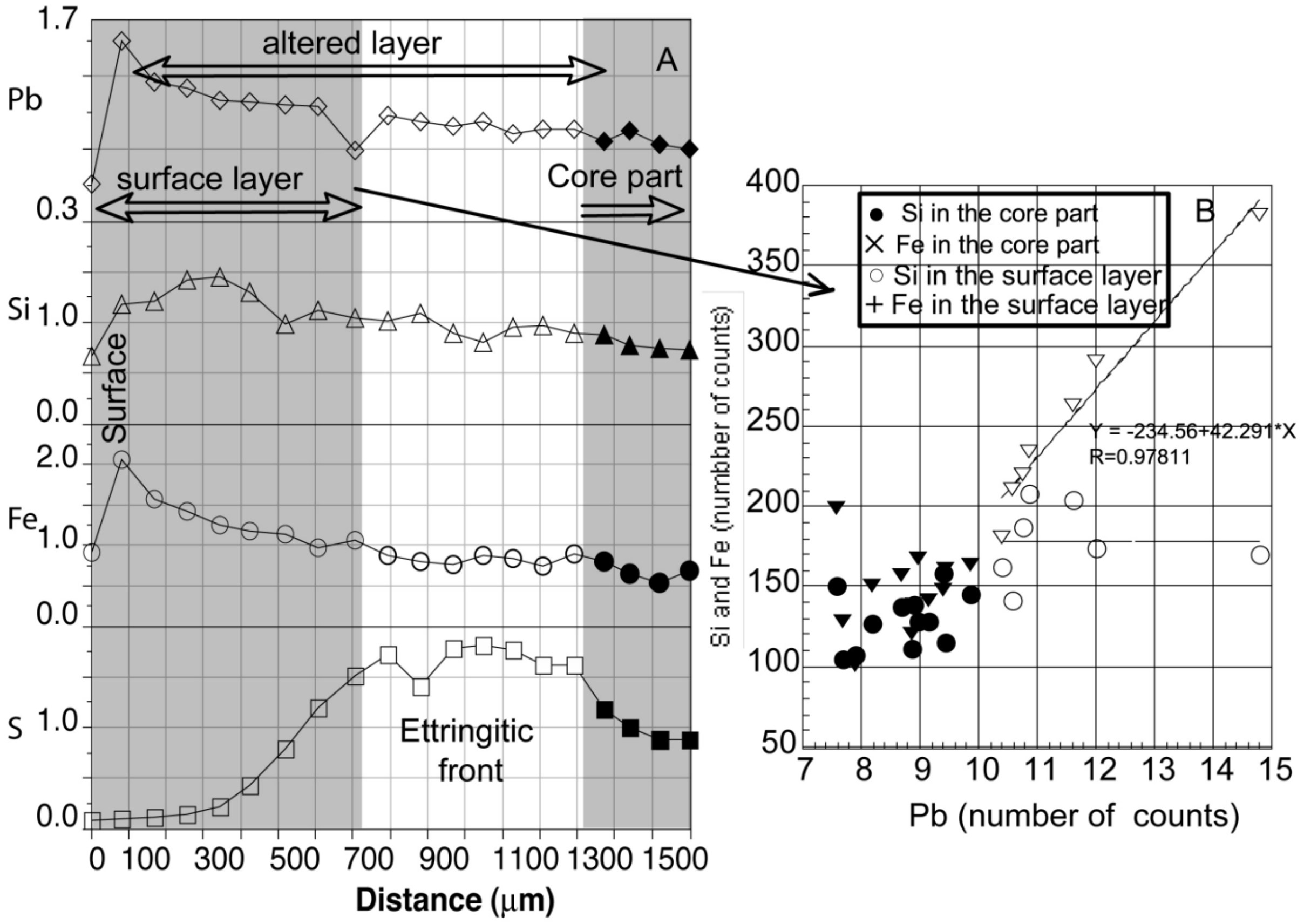

Figure 3. A) μ-XRF profile of Pb, Si, Fe and S from the unaltered core to the altered layer of the leached Portland cement. B) Correlation between Pb and Si and Fe in the unaltered core and altered layer (0 to 700 μm)

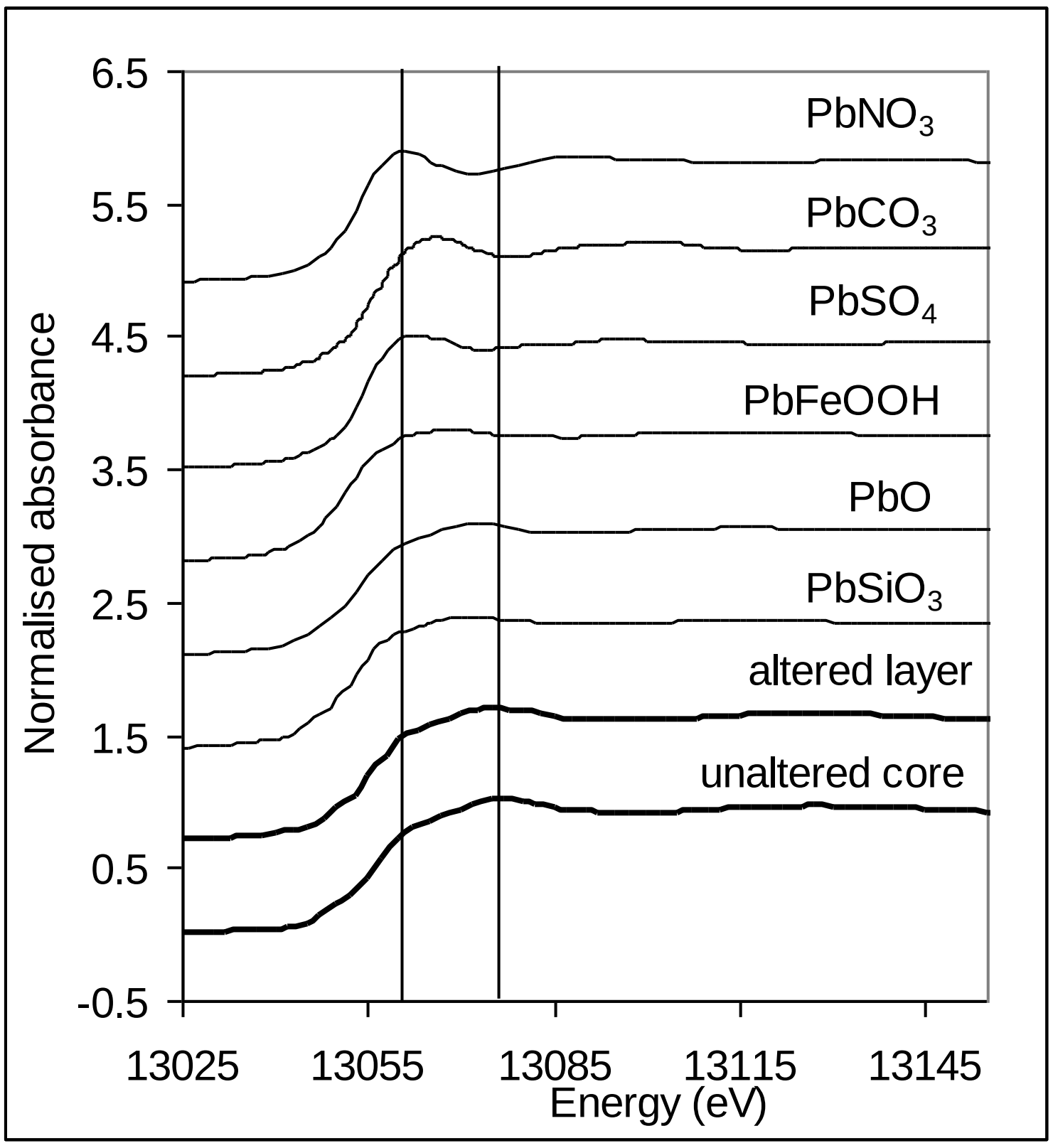

Figure 4. Comparison of Pb $L_{III}$ XANES spectra of unaltered core, altered layer, $PbNO_3$, $PbCO_3$, $PbSO_4$, PbFeOOH, PbO and $PbSiO_3$.

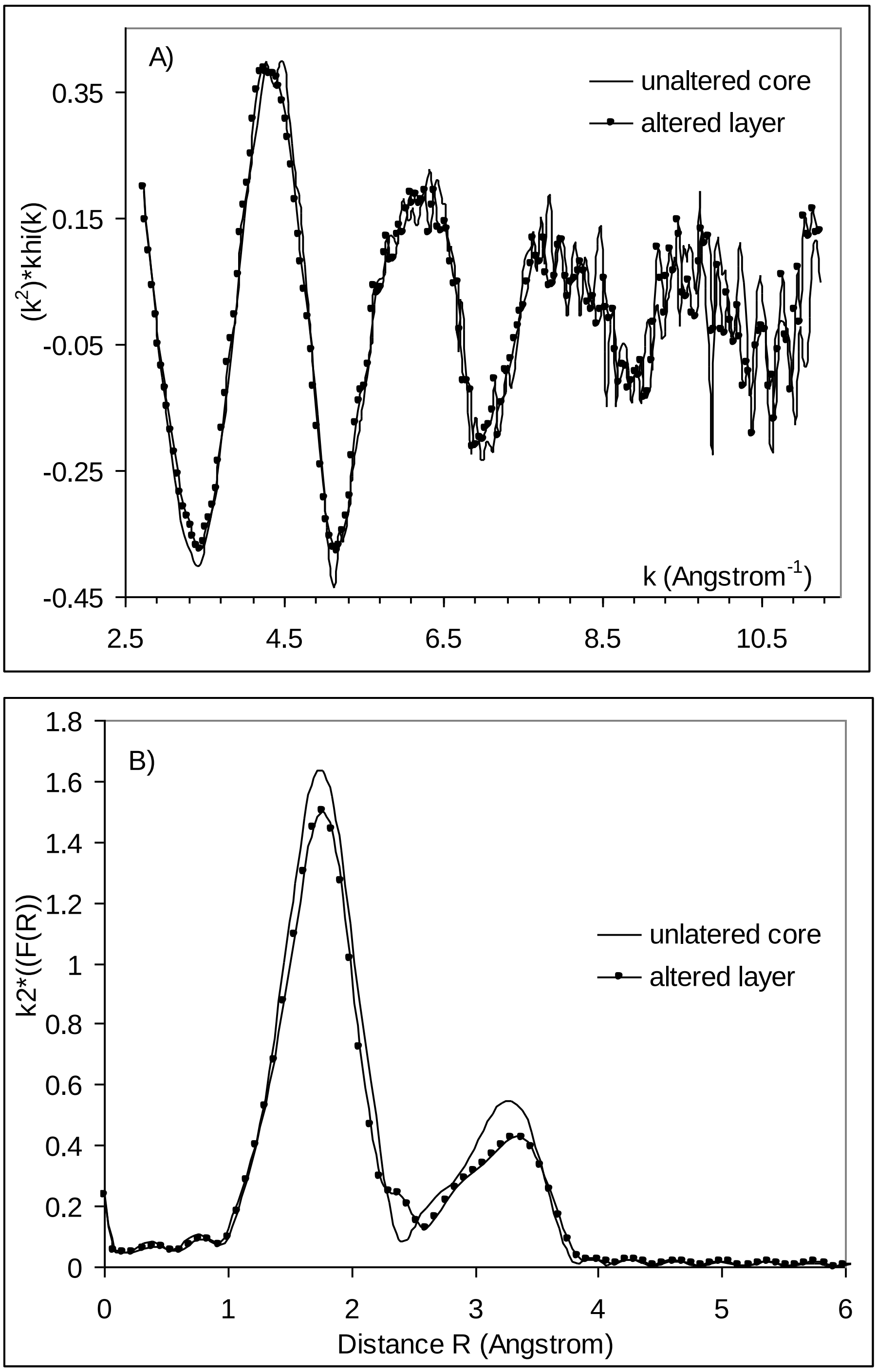

Figure 5. Comparison between Pb $L_{III}$ edge A) EXAFS and B) RDF spectra in the unaltered core and altered layer of the leached Portland cement.

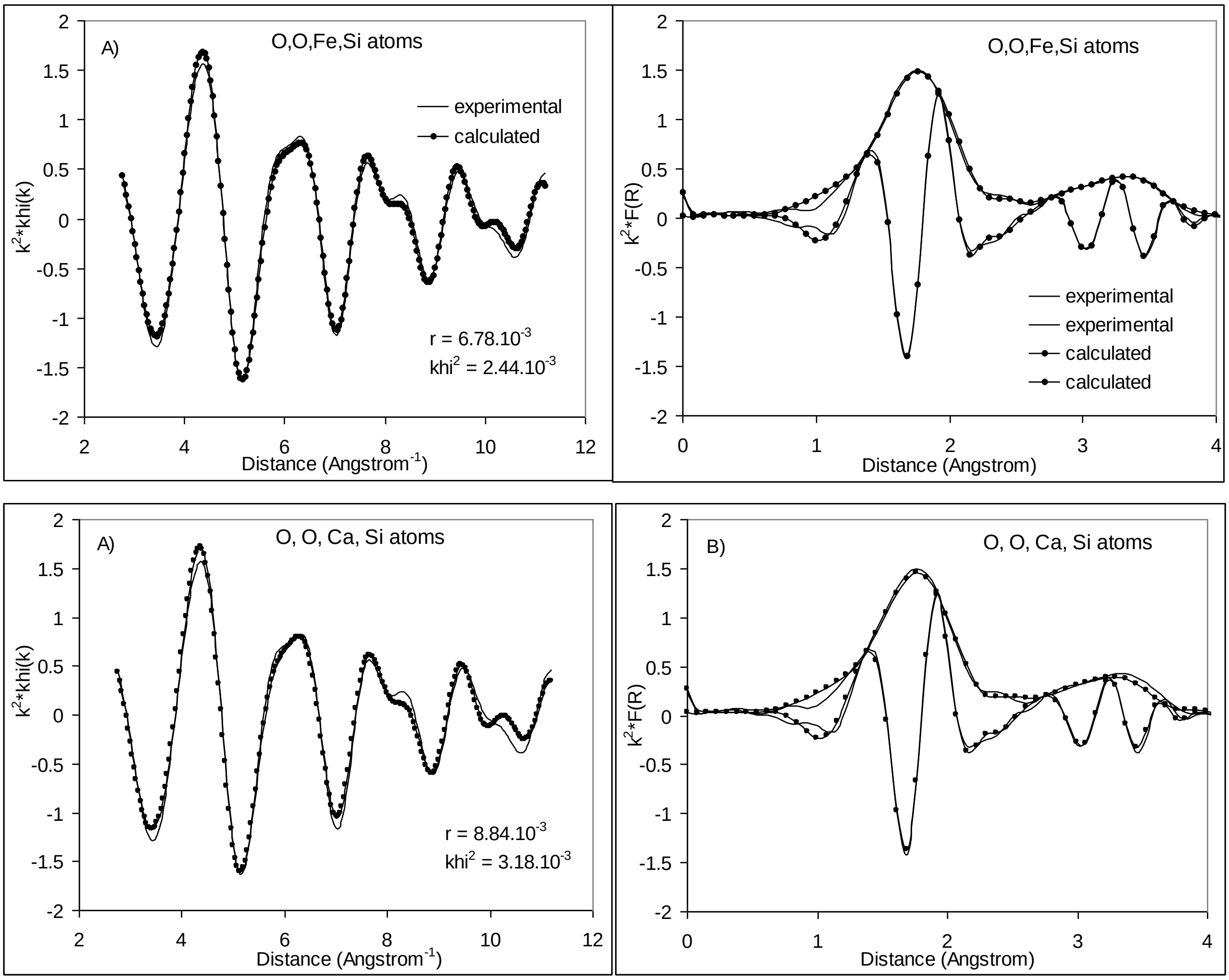

Figure 6. Simulation of Pb $L_{III}$ EXAFS spectra of the altered layer and comparison between A) EXAFS and B) RDF (modulus and imaginary part) for the 2 solutions (O, O, Si, Ca and O, O, Si, Fe atoms in Pb atomic environment).

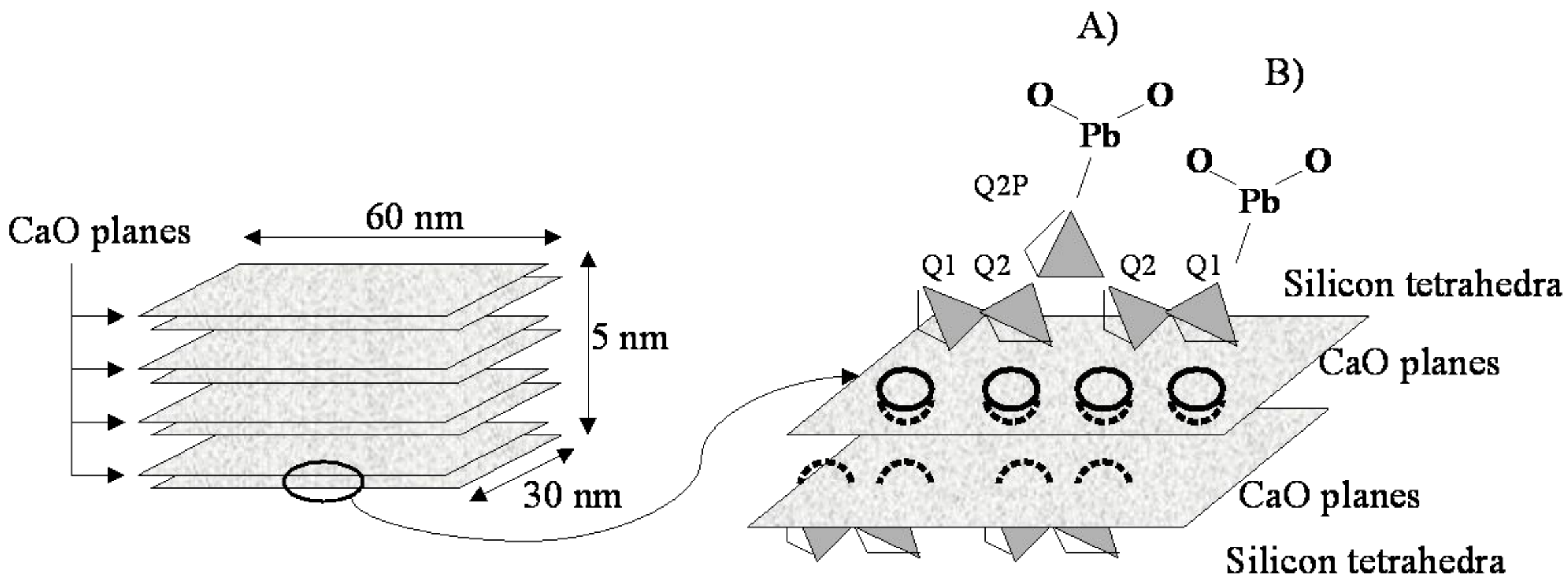

Figure 7. Synthesis of EXAFS results: Pb retention sites through C-S-H nanostructure in the unaltered core and altered layer of C-S-H, according to C-S-H model described in [16]. A) Case of Pb(II) linked with $Q_{2P}$ silicon tetrahedra, B)Case of Pb(II) linked with $Q_1$ silicon tetrahedra.

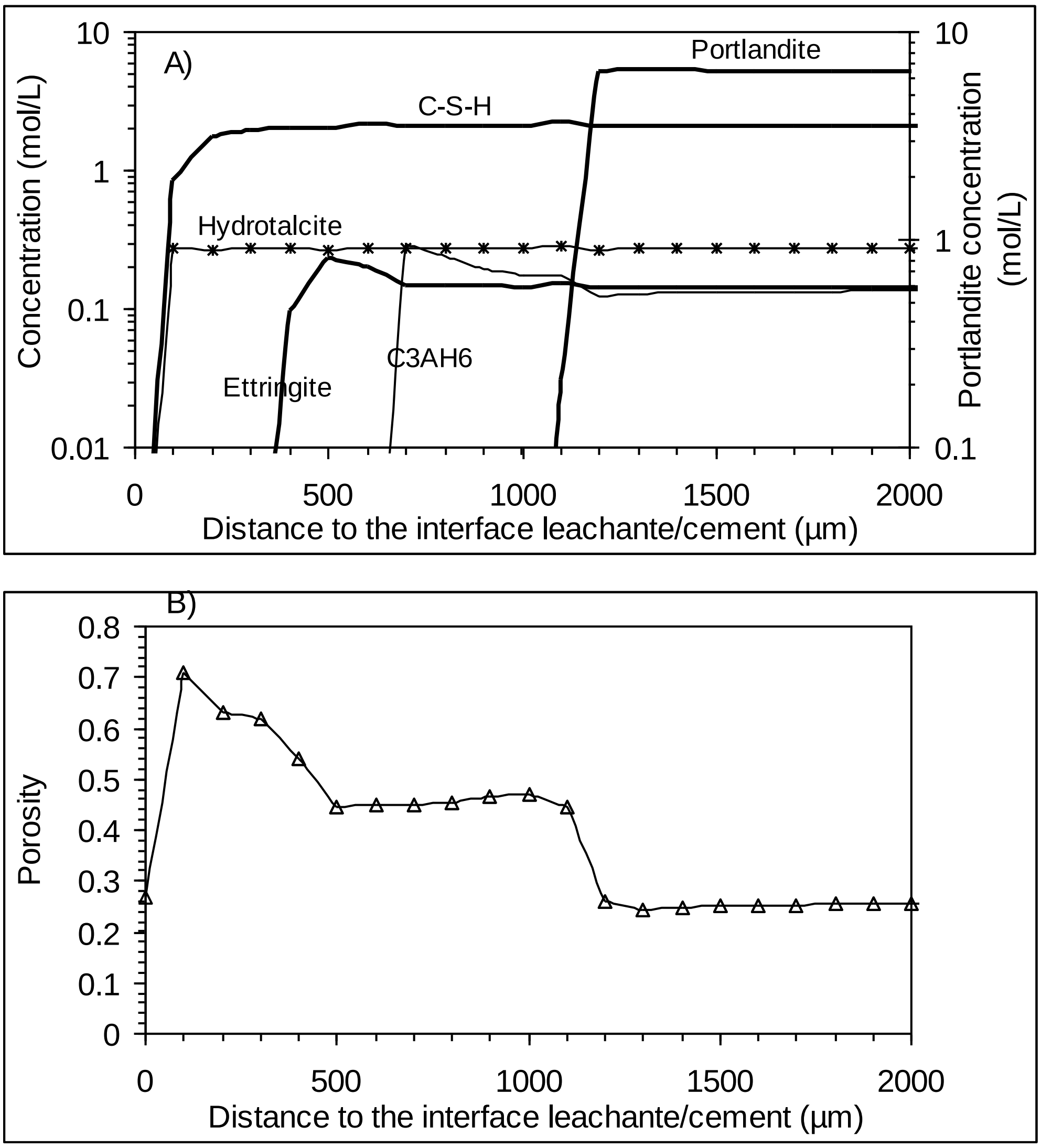

Figure 8. A) Evolution of cement mineralogy after 50 days of leaching simulation with CHESS-HYTEC code. On the left, the y axe is referred to the concentration of ettringite, hydrotalcite, C-S-H and $C_3AH_6$, in mol/L of porous volume, and on the right the y axe is portlandite concentration in mol/L of porous volume. B) Evolution of cement porosity after 50 days of leaching simulation with CHESS/HYTEC code.

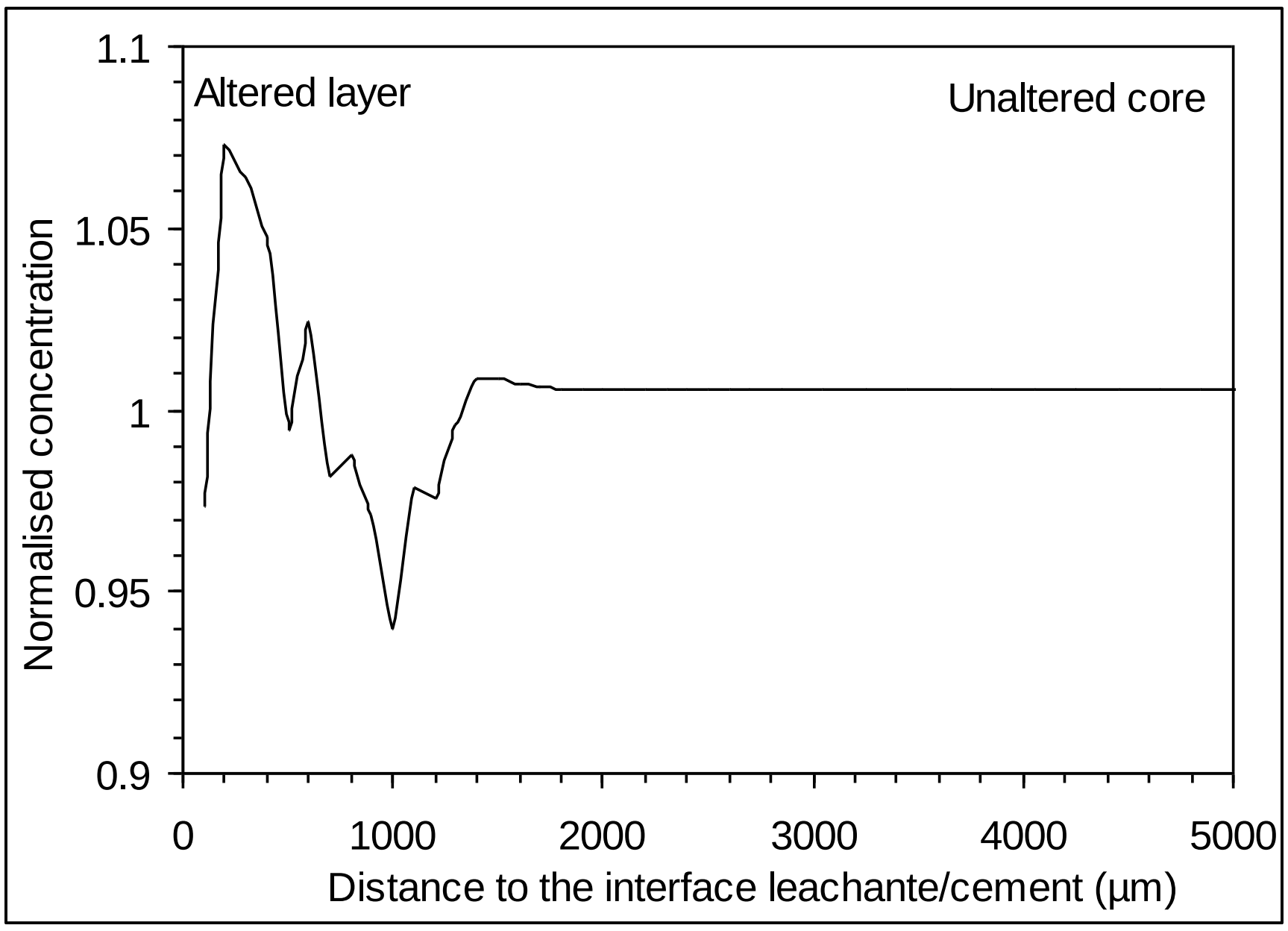

Figure 9. Profile of Pb behaviour in cement matrix after 50 days of leaching simulation with CHESS/HYTEC code.